\documentclass[print]{aa} 

\usepackage{graphicx}
\usepackage{natbib}
\newcommand{\be}{\begin{eqnarray}}
\newcommand{\ee}{\end{eqnarray}}

\newcommand {\nbodypp}{\textsc{\mbox{nbody6\raise.4ex\hbox{\tiny++}}}}
\newcommand {\COri} {\mbox{$\theta^1{\rm{C}}\:{\rm{Ori}}$}}

\newcommand {\Msun} {\mbox{M$_{\odot}$}}

\begin{document}

\title{Gravitational Instabilities induced by Cluster Environment? - The encounter-induced angular momentum transfer in discs}
\author{S.~Pfalzner \and C.~Olczak} 
\institute{I. Physikalisches Institut, University of Cologne, Z\"ulpicher Str. 77, 50937 Cologne, Germany}
\date{}

\abstract
{}
{The aim of this work is to understand to what extend gravitational interactions between the stars 
in high-density young stellar clusters, like the Orion Nebula Cluster (ONC), change the
angular momentum in their protoplanetary discs.}
{Two types of simulations were combined ---  N-body simulations of 
the dynamics of the stars in the ONC, and angular momentum loss results from simulations 
of star-disc encounters.}
{It is shown that in a star-disc encounter the angular momentum loss is usually larger than the 
mass loss, so that the disc remnant has a lower specific angular momentum. Assuming an age of 
1-2 Myr for the ONC, the disc angular momentum in the higher density region of the Trapezium is reduced
by 15-20\% on average. Encounters therefore play an important part in the angular momentum transport in these
central regions but are not the dominant process. More importantly, even in the outer cluster regions
the angular momentum loss is on average  3-5\%.  
Here it is shown that a 3-5\% loss in angular momentum might be enough to trigger gravitational 
instabilities even in low-mass discs - a possible prerequisite for the formation of planetary systems.}
{}

\titlerunning{Cluster as trigger for gravitational instabilities}
\authorrunning{Pfalzner et al.}

\maketitle

\section{Introduction}

Discs are seen as a prerequisite for the formation of planetary systems.
There is increasing observational evidence that most, if not all, stars are initially surrounded
by discs. For example, Lada et al. (2000) found that 80-85\% of all stars in the Trapezium are 
surrounded by discs.


One visually compelling feature in the numerical study of these discs is the development of 
spiral arms either in self-gravitating high-mass discs or as an effect of encounters 
\cite{pfalzner:03}. In a few cases where discs have been observed with a sufficiently high 
spatial resolution, the existence of such spiral structures have been observationally confirmed  
\citep[for example]{clampin:03,corder:05,lin:06}. Spiral arms are always connected to mass and
angular momentum transport. Our study of the importance of the encounter-induced
angular momentum transport in an cluster environment has two motivations: angular 
momentum transport (a) in the late stages of accretion in star formation and (b)in spiral arm
formation as prerequisite for gravitational instabilities and fragmentation in planet
formation scenarios. 

The former is connected to the problem that the typical observed angular momentum of 
the cloud cores is so large \citep{bodenheimer:araa95} that the star would be 
spinning  at velocities higher than its break-up velocity if no means for angular momentum
loss would be present. One way out is angular momentum transfer to the disc 
and then from the inner disc regions outwards. Many different mechanisms have been proposed 
for the latter \citep[see][and references therein]{larson:mnras02}.
%
%
Although often discarded as probably not important, encounter-induced angular
momentum transport is still often cited as possible mechanism. To clarify this point
we study disc angular momentum loss in a cluster environment for the first
time in a quantitative, systematic manner. 

The second aim of this investigation is to achieve a better understanding of how 
the cluster environment influences the development of spiral structures in the discs
through encounter-triggered angular momentum transport. This might be of special importance
as thus such spiral structures are a prerequisite in the theory of planet formation through 
gravitational instabilities. 

The Orion Nebula Cluster (ONC) was chosen as the model cluster for these investigations
because 
it is one of the best-studied regions in our galaxy, and observational constraints 
significantly reduce the modelling parameters. In addition, its high density suggests 
that stellar encounters might be relevant for the evolution of circumstellar discs.  
This is supported by recent investigations by Olczak et al. (2006) indicating that
the effect of encounters has - at least for the disc mass loss - previously been 
underestimated.

The investigation involves three steps --- i) the determination of  the encounter 
parameters in the ONC dynamics, ii) the angular momentum loss in isolated encounters, and 
iii) the combination of both results to investigate
the angular momentum loss of the discs in the cluster.

For the first step we use the cluster simulations described in Olczak et al. (2006) and 
star-disc simulations in Pfalzner et al. (2006)
to determine a parameter list for the encounters for a model ONC in equilibrium and expansion,
i.e. virial ratios of $Q_{vir}$=0.5 and  $Q_{vir}$=1. The results are summarized in Section 2.

This is followed by an investigation of the angular momentum loss in star-disc encounters 
in Section~3. There have been earlier investigations which looked at angular momentum transport in 
isolated star-disc encounters \citep{ostriker:apj94,larwood:97,hall:mnras96,pfalzner:apj04}
predominantly in prograde, coplanar encounters for low-mass discs covering only a small
parameter range. Under these limitations simple N-body simulations suffice, 
and hydrodynamic effects and self-gravity within the disc can be largely neglected. 
Following this approach one has to keep in mind that the results have to be considered
as upper limits (see discussion in Olczak et al. (2006) for the disc mass loss).
We extended the parameter range covered in Pfalzner et al. (2005a) to larger mass ratios 
between the interacting stars and penetrating encounters  to cover all situations necessary 
when modelling the ONC. In this work we neglect photo-evaporation as disc dispersal mechanism 
\citep{scally:mnras01}, but this should be included in a future investigation.

In Section~4 the results of Section~2 and 3 are combined to determine the disc angular momentum 
as a function of time. In Section~5 it is shown how the angular momentum mass loss is 
influenced by the assumptions made and a discussion is given of how this angular momentum loss 
can trigger gravitational instabilities.

\section{Model and Cluster simulations}


We combine simulations of the dynamics of the ONC determining the interaction parameters 
of close encounters between stars in the cluster \cite{olczak:apj06} and with results of 
isolated star-disc encounter simulations described in \cite{pfalzner:aa06}. In contrast to these 
previous studies, the emphasis in the current investigation is not on the mass but the angular momentum 
loss of the discs. 

The dynamical model of the ONC presented here contains only stellar components neglecting 
gas and the potential of the background molecular cloud OMC~1.
The cluster models were set up with a spherical density distribution $\rho(r)\propto r^{-2}$,
by  randomly generating the masses according to the mass function given by Kroupa et al. (1993) 
in a range $50 \Msun \ge M^* \ge 0.08 \Msun$. \COri\ was directly assigned a mass of 50 \Msun \ 
and placed at the cluster centre. The velocity distribution of the stars was 
assumed to be Maxwellian.

The ONC cluster simulations were performed with \nbodypp\ \citep{spurzem:mnras02} and modelled
for 13\,Myr -- the assumed lifetime of \COri. 
The quality of the dynamical models was judged by comparing to observational data at 
1-2\,Myr, marking the range of the mean ONC age. For more details of the selection 
process, see Olczak et al. (2006). The quantities of interest were: number of stars, 
half-mass radius, number densities, velocity dispersion and projected density profile. 
Here we describe mainly the case of the ONC being in virial equilibrium, in 
Section~\ref{sec:angONC} the changes one can expect for an expanding cluster will be briefly considered. 

Assuming that only two-body encounters occur and that higher-order encounters are negligible, 
the effect of an encounter can be investigated by considering it to be isolated from the 
rest of the cluster. During the course of the simulation the information 
of all perturbing events of each stellar disc, i.e. both masses, the relative 
velocity and the eccentricity are recorded. However, the latter is not applied as such
later on. To limit the parameter space only the angular momentum loss of parabolic encounters 
is used. The eccentricity values are utilized to estimate the error introduced
by such a simplification.
\section{Angular momentum loss in star-disc encounters}
\label{sec:star-disc}
This part of the investigation is mainly based on the earlier work by 
Pfalzner et al. (2005a) where a parameter study  was performed for a star of mass 
$M_1^*=1\,\Msun$ surrounded by a disc perturbed 
by the fly-by of a second star of mass $M_2^*$.
When simulating isolated encounter events, the observational evidence that most discs in 
the ONC are of low mass $m_{d}$, i.e. $m_{d}/M^* \ll$ 0.1, reduces the complexity of 
the numerical approach in several ways: low-mass discs do not significantly influence the 
encounter orbit and self-gravitation can be neglected and the results are 
scalable to other star masses.  The effect of viscosity is in this case negligible too, as
the viscous timescale is so much longer than the timescale of the encounter.
Another simplification is that only coplanar, prograde 
encounters were investigated. According to studies on inclined and retrograde star-disc
encounters \citep{ostriker:apj94,heller:apj95,hall:mnras96,pfalzner:aa05}, due to 
these approximations the results can only be interpreted as an upper limit for the angular 
momentum loss. We will dicuss this in more detail at the end of this section.

The disc is assumed to extend to $r_d=100$\,AU and the surface density initially has 
a $1/r$-dependence. The angular momentum loss induced by the fly-by of 
a star on such a star-disc system depends on various parameters specific to the 
stars and discs involved and the encounter orbit. The above mentioned earlier study 
\citep{pfalzner:aa05} was 
limited to non-penetrating encounters with star mass ratios in the interval 
$0.1<M^*_2/M^*_1<5$. Here, the parameter study is extended to the higher mass ratios of up to 
$M_2^*/M_1^*=50.0\,\Msun/0.08\,\Msun$ which  
occur in the ONC simulations. The additional fractional angular momentum loss (AML) 
values can be found in Table~\ref{table:jloss}
available with the electronic form of this article. The improved fit function 
for   $\left(\Delta J/J\right)_{total}$ is
\begin{eqnarray}
    \left(\frac{\Delta J}{J}\right)_{total} & = &  
    1.02 \left(\frac{M_2^*}{M_1^*+M_2^*}\right)^{0.5r_p}\nonumber\\
    & & \times \exp\left[-\sqrt \frac{M_1^*\left(r_p-0.7r_p^{0.5}\right)^{3}}{M_2^*}\right]
 \label{eqn:Fit_total}
\end{eqnarray}
where $r_p = r_{peri}/r_{disc}$ is the periastron separation $r_{peri}$ in units of the disc radius $r_{disc}$.
%
%
\begin{figure}
\resizebox{\hsize}{!}{\includegraphics[angle=-90]{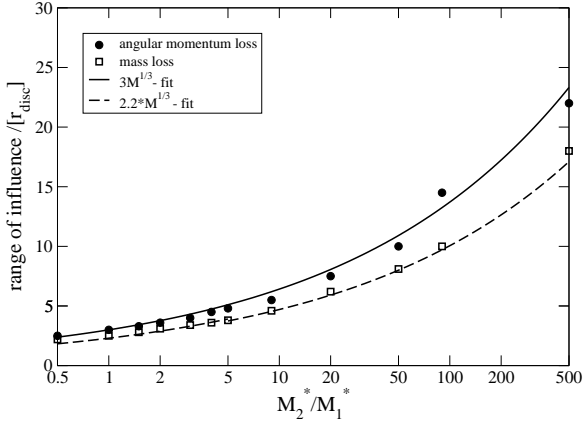}}
\caption{Interaction range for mass and AML as a function of the mass ratio of the interacting stars in units of the disc radius.}
\label{fig:jmass}
\end{figure}

The most important results for the cluster are that 
\begin{enumerate}
\item the AML is always larger than the relative mass loss,
\item the angular momentum of the disc 
is affected by much more distant encounters than the disc mass. 
\end{enumerate}

The latter is illustrated by Fig.~\ref{fig:jmass}, where the relative distance 
$r_{peri}/r_{disc}$ at which the angular momentum is changed by at least 3\% (the average 
error value) is shown as a function of the 
mass ratio $M_2^*/M_1^*$. Given a sufficiently large perturber mass, 
even encounters as distant as 20 times the disc radius can reduce the angular 
momentum in the disc by 10\% or more without changing the disc mass at all.

\begin{figure}
\resizebox{\hsize}{!}{\includegraphics[angle=0]{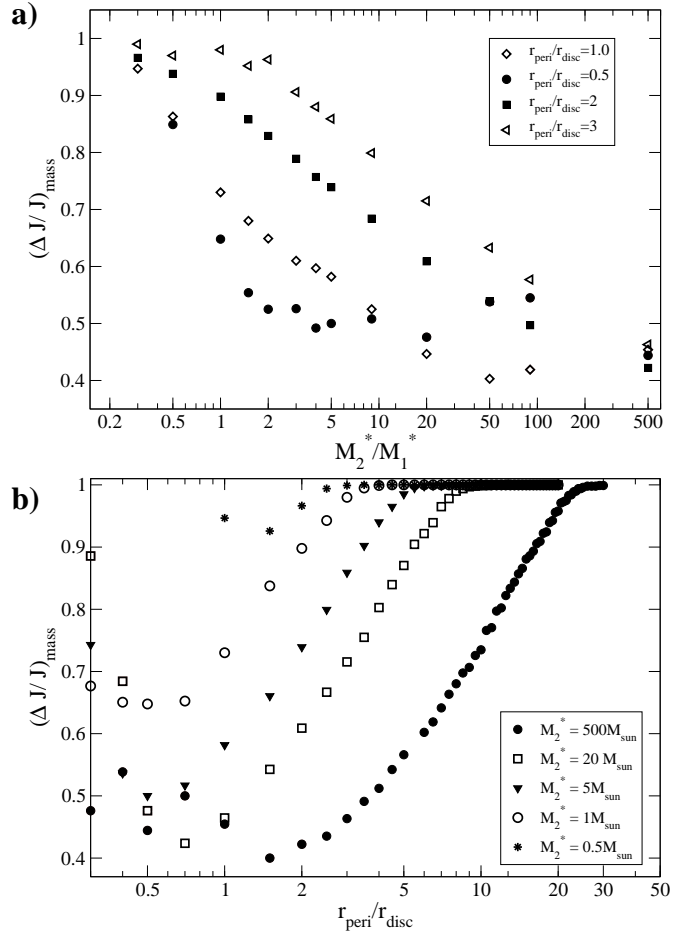}}
\caption{Specific angular momentum after an coplanar, prograde parabolic encounter
compared to that of the initial disc shown as a function of a) the perturber mass
$M_2^*$ and b) the encounter periastron in units of the disc size.}
\label{fig:relmass}
\end{figure}

A considerable part of the angular momentum is carried away with the matter that becomes 
unbound. However, what one is really interested in is to what degree the angular momentum 
per particle is lower in the remaining disc --- only this can eventually facilitate accretion 
of matter onto the star. A measure of this specific angular momentum can be defined by 
\be
\left(\Delta J/J\right)_{mass} := 
\left(1-\displaystyle {\frac{\Delta J}{J}}\right) \mbox{\huge/}
\left(1-\displaystyle {\frac{\Delta m}{m}}\right).
\label{eq:j_total}
\ee
Fig.~\ref{fig:relmass}a and b show the specific angular momentum obtained
by using Table~\ref{table:jloss} in this paper and Table 3 in Olczak et al. (2006)
as a function of $M_2^*/M_1^*$ and $r_{p}$, respectively. 
\footnote
{Note,  fit formula \ref{eqn:Fit_total} here and Eq.~4 in Olczak et al. (2006) are not 
of high enough accuracy to determine $\left(J_{enc}/J\right)_{mass}$ in 
Eq.~\ref{eq:j_total} in the entire parameter range required for the ONC.}
For large mass ratios there exists
an lower limit of 0.4 for $\left(J_{enc}/J\right)_{mass}$ for all considered 
periastra. This means that in a single encounter, however strong, the specific 
angular momentum
can only be reduced to $\sim$40\% of that in the initial
disc. However, such a change of angular momentum is achieved over a wide range of 
interaction parameters, so that in repeated encounters the angular momentum of the 
remaining disc could nevertheless be considerably reduced.

Since the cluster consists of a wide spectrum of star masses, these 
simulation results obtained for $M_1^*=1\,\Msun$ are generalized by scaling the disc 
radius according to
\begin{eqnarray}
    r_{disc}
    =r_{disc}(1\,\Msun)\sqrt{M_1^*[\Msun]}
    \label{eqn:rdisk}
\end{eqnarray}
In the cluster simulations $r_{disc}(1\,\Msun)$ was assumed to be 150AU.
Scaling the disc size with the star mass seems intuitively right, but 
observations are not so unambiguous: Vicente \& Alves (2005) found a correlation between disc 
diameters and stellar masses using a sample of proplyds from 
Luhman et al. (2000), but not for the data from Hillenbrand (1997). 
However, the present Trapezium is probably not in its primordial state
as various disc destruction processes have most likely altered the disc sizes considerably.


\begin{figure}
\resizebox{\hsize}{!}{\includegraphics[angle=0]{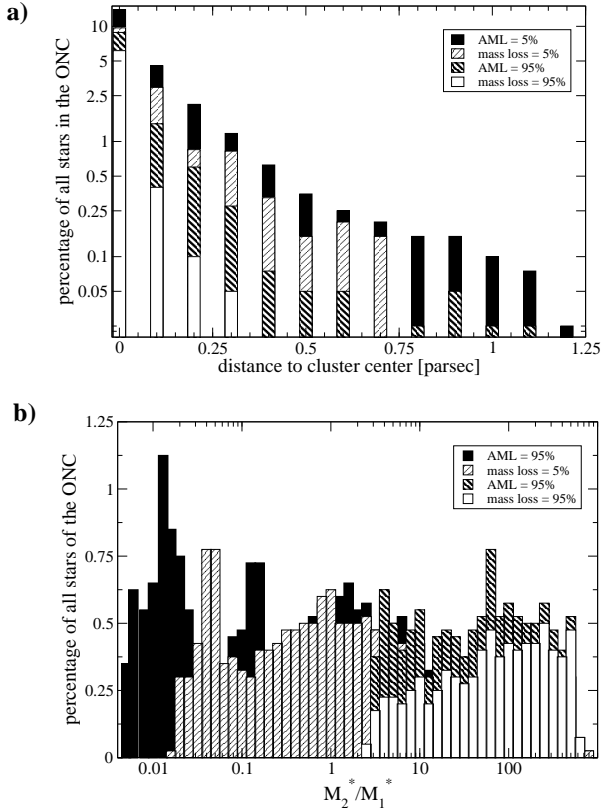}}
\caption{a) Percentage of stars in bins with a 5\% (lowest considered bin) or 95\% 
(highest considered bin) AML in the 
disc as a function of distance to the cluster center at 2 Myr averaged over 
20 simulations.  b) shows the same as a function of the mass ratio $M_2^*/M_1^*$
of the two interacting stars. For comparison in both figures the same is shown 
for the mass loss, too.}
\label{fig:ang_r}
\end{figure}

As mentioned earlier the encounter results summarized by Eq.~\ref{eq:j_total}, 
Fig.~\ref{fig:relmass} and Table~\ref{table:jloss} represent upper limits to the angular momentum 
loss induced by encounters. The reason is that they are obtained for prograde, coplanar and
parabolic encounters only. In addition, it was assumed that only one of the stars was 
surrounded by a disc. The latter is somewhat contradicting - assuming that
each star is initially surrounded by a disc but at the same time considering an encounter
with a disc-less star. 
There are two reasons why this was done: first, encounters where both stars are surrounded 
by discs are less well investigated and second Pfalzner et al. (2005b) showed that the
star-disc results can be generalized to disc-disc encounters as long as there
is no mass exchange between the discs. In the case of a mass exchange the discs
can be to some extent replenished with material of very low angular momentum
so that the  specific angular momentum loss (SAML) would be underestimated.

Most encounters in the cluster simulations are not parabolic but hyperbolic. In such hyperbolic 
encounters the angular momentum loss is lower because the disturber is not long enough in the
vicinity of the star-disc system to remove angular momentum in an efficient way. 
However, considering only the stars that lose more than 90\% of their disc mass, the 
eccentricity distribution has a maximum at $\epsilon \approx 3$.
Pfalzner et al.(2005a) showed that for $M_2^*=1\,\Msun$ the AML in an
$\epsilon = 3$ encounter is about 55\% of that of a parabolic encounter.

In a cluster that is not highly flattened it seems rather unlikely that the encounter 
planes are aligned to a high degree. Therefore one would expect most encounters to be 
non-coplanar. Pfalzner et al. (2005a) showed that, as long as the inclination is not larger than  
45$^\circ$ the AML in the encounter is only slightly ($<$ 10\%) reduced in comparison 
to a coplanar 
encounter. If however the orientation is completely random and a 90$^\circ$-encounter the
most likely encounter scenario, the mass loss could be significantly reduced, a point which would need 
further investigation.

It is often argued that consecutive encounters would not lead 
to any significant additional momentum transport because the outer edge of the
disc is truncated by the first encounter - the part that is mainly involved 
in the interaction. However, Pfalzner et al. (2004) demonstrated that a second encounter can 
indeed lead to the same {\it relative} AML than in the first 
encounter.  This implies that a succession of distant encounters might well be able to 
transport a higher amount of angular momentum outwards than previously thought
\citep[see][too]{moeckel:06}. In future studies, it should be investigated whether this holds 
for the entire parameter space covered here.  
 
\begin{figure}
\resizebox{\hsize}{!}{\includegraphics[angle=0]{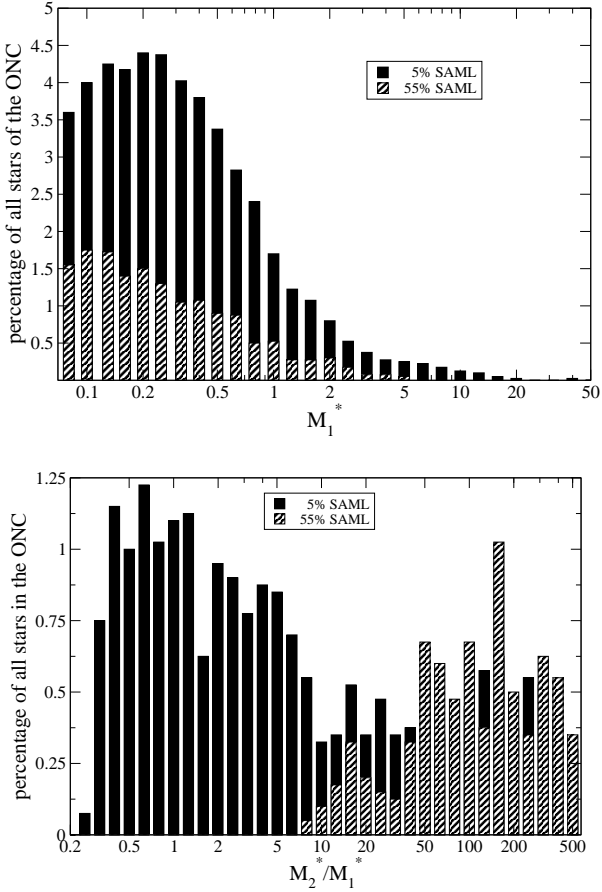}}
\caption{Percentage of stars with a SAML as defined by 
Eq.~\ref{eq:j_total} as function of the stellar mass averaged over 10 simulations.
The bin for 5\% and now 55\% (as the highest obtainable SAML is $\sim$ 60\%) loss are
shown.}
\label{fig:ang_mass_para}
\end{figure}

\section{Angular Momentum Loss in the ONC}
\label{sec:angONC}

%
%

In the following the cluster results are combined with those of 
Section~\ref{sec:star-disc} to investigate the
encounter-induced angular momentum loss of the discs in the ONC.
First we determine the AML using Eq.~\ref{eqn:Fit_total}. 
The fact that the angular momentum loss is usually larger than the mass loss, 
is reflected by the cluster simulations in several ways:
As Fig.~\ref{fig:ang_r}a) shows, the number of stars with, for example, more than 5\% or 95\% 
relative angular momentum loss in the disc is somewhat higher than the number of discs
with a more than 5\% or 95\%  mass loss. Due to the lower stellar number density in the outer 
cluster regions both the mass loss and the AML decrease with distance from the
cluster center. However, the AML extends to more 
distant cluster regions --- for the 5\% case the mass loss extends to 
$\sim$ 0.75 pc , whereas the AML reaches to $\sim$ 1.25 pc.  
For high losses (95\%) the mass loss extends only to 0.3-0.4pc, but the angular momentum 
to $\sim$ 1.1pc. Fig.~\ref{fig:ang_r}b demonstrates that lower relative perturber masses 
$M_2^*/M_1^*$ are required to remove at least 5\% of the angular momentum than mass from the disc.


\begin{figure}
\resizebox{\hsize}{!}{\includegraphics[angle=-90]{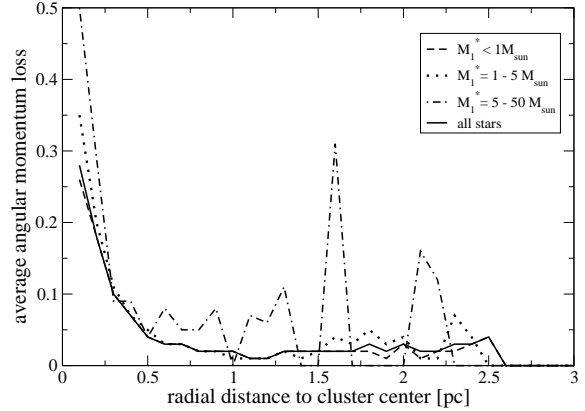}}
\caption{SAML 
as a function of the radial distance to the cluster center averaged over 20 simulations.}
\label{fig:angradmean}
\end{figure}

The dependence on the distance from the cluster center of the SAML is very similar to that of
the  AML, i.e. whereas beyond 0.3pc no mass loss occurs, AML and SAML happen as far out as 
$\sim$ 1.25 pc.  However, some differences nevertheless occur. They are due to the fact
most massive stars have basically lost all their mass and are not considered anymore for
this mass-corrected value. In addition, the dependence on $M_2^*/M_1^*$ is moved to higher 
$M_2^*/M_1^*$-values (see Fig.~\ref{fig:ang_mass_para}b). This 
might indicate that low-mass encounters can remove only mass with
angular momentum, whereas high-mass encounters can do the opposite.

Fig.~\ref{fig:angradmean} shows that the SAML is reduced most in the central region of 
the cluster (by $\sim$ 15-20\%), but throughout the cluster there is SAML even if only on a
3-5\% per cent level in the outer cluster areas. 

\begin{figure}
\resizebox{\hsize}{!}{\includegraphics[angle=-90]{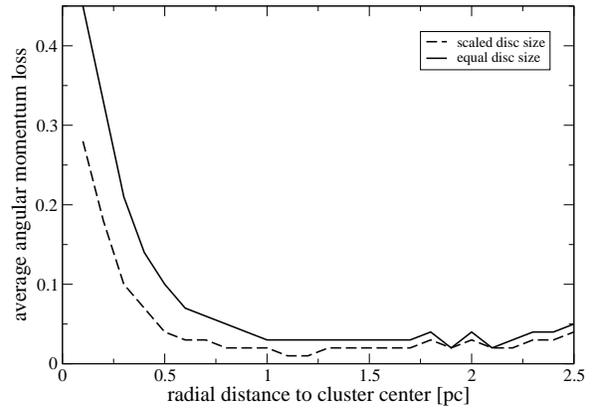}}
\caption{SAML for equal and scaled disc size
as a function of the radial distance to the cluster center.}
\label{fig:angradeq}
\end{figure}

As mentioned in Section 3 it is not clear what the primordial disc-size distribution in 
a cluster is. If we assume that the disc size does not scale with the disc mass, but 
assume initially equal disc sizes (150AU), the SAML is larger at all distances 
(see Fig.~\ref{fig:angradeq}). This is most pronounced in the Trapezium ($\sim$ 30-40\%), 
but even further outside the SAML is increased by $\sim$ 1\%.
The reason is that the majority of stars have masses $<$1$\Msun$. Since an equal disc size
for all stars is equivalent to a larger disc size for them, this results in a higher
SAML.

\begin{figure}
\resizebox{\hsize}{!}{\includegraphics[angle=-90]{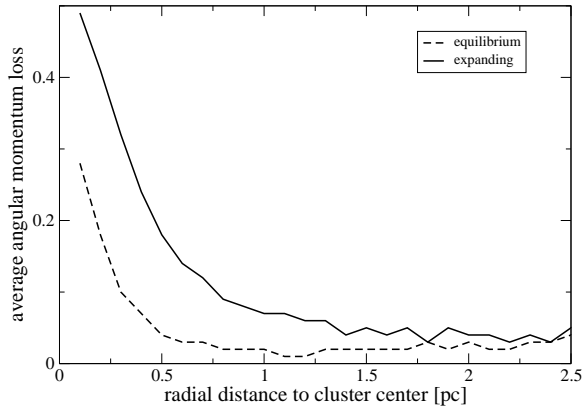}}
\caption{SAML as function of radial distance to cluster center. Comparison between
cluster in equilibrium and an expanding cluster for scaled disc size.}
\label{fig:angrad_exp}
\end{figure}

As the velocity dispersion of the ONC is not well determined, the next point to investigate
is how the SAML  changes if we consider the cluster to be expanding.
Testing the $Q_{vir}$=1 case, the angular momentum reductions turns out to be larger 
in the entire ONC
with the largest differences in the Trapezium region (see Fig~\ref{fig:angrad_exp}).

\begin{figure}
\resizebox{\hsize}{!}{\includegraphics[angle=-90]{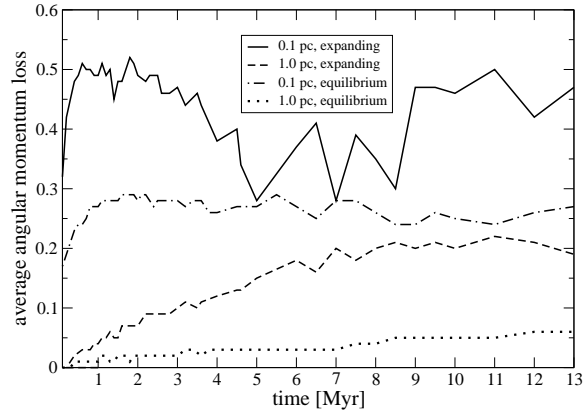}}
\caption{SAML as function of time for an expanding cluster at
0.1pc and 1.0pc.}
\label{fig:angtime_exp2}
\end{figure}

Fig.~\ref{fig:angtime_exp2} shows that the difference in SAML happens
early on in the cluster development. Here the density in the expanding cluster
is initially higher in the center leading to more and stronger encounters.

\section{Discussion and Conclusions}
\label{sec:discussion}

Combining simulations that follow the path of the stars in the cluster with 
angular momentum loss results obtained from simulation of isolated encounters,
the encounter-induced angular momentum loss has been investigated
for the ONC. 
The main outcome of these simulations is that for most stars in the ONC, 
encounters reduce the angular momentum on average only  by 3-5\% in 
the remaining disc compared to the initial disc. However,
this kind of reduction happens even in the outer parts of the ONC, so that
the disc angular momentum is affected by encounters throughout the entire 
cluster. If one considers only the central dense Trapezium region 
the SAML increases to the 10-30\% level.  
This means that although encounters in the cluster environment do 
lead to angular momentum loss in the disc, it is not the 
predominant process for angular momentum transfer. 

\begin{figure}
\resizebox{\hsize}{!}{\includegraphics[angle=0]{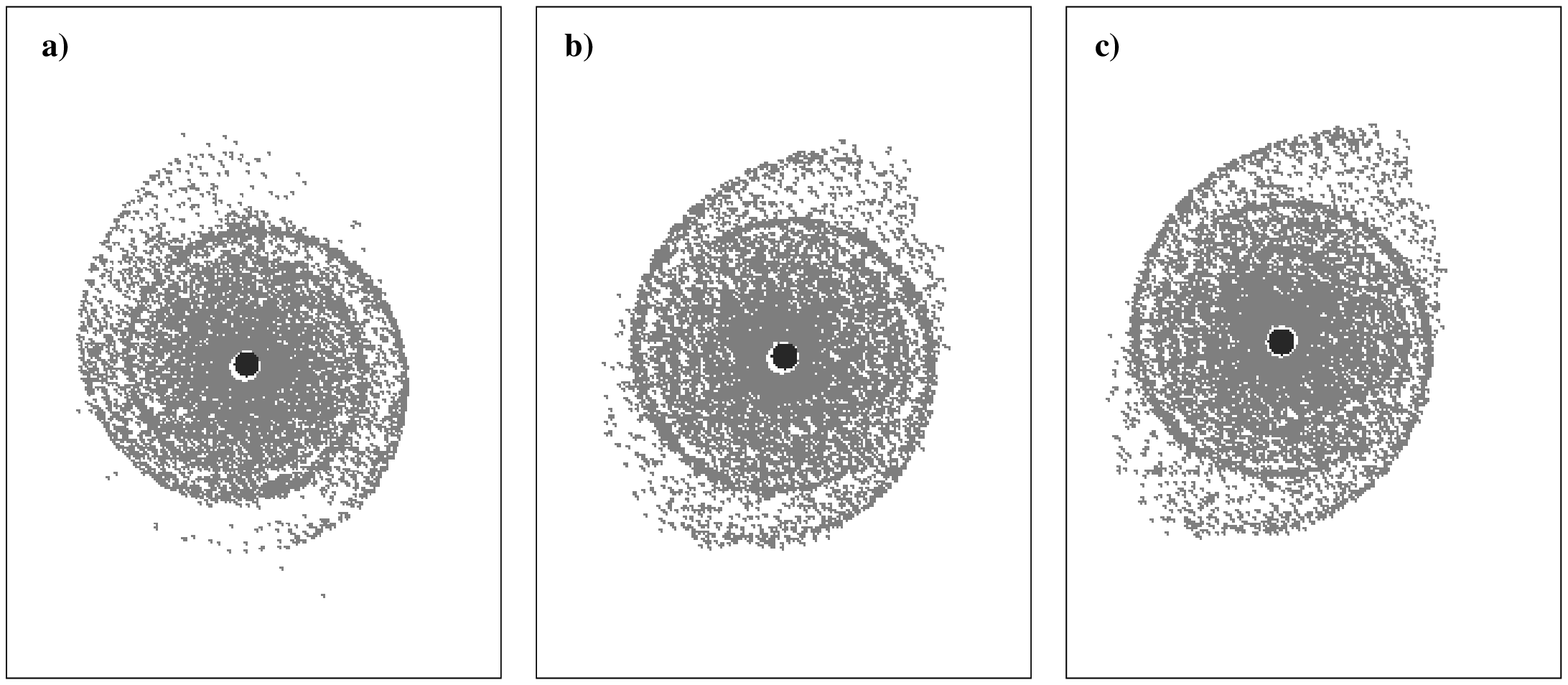}}
\caption{Examples of encounters with a (SAML) of 3-5\%. a) shows an encounter where
$M_2^*$= 0.1 $M_{\sun}$ and $r_p$ = 2.0, b)  an encounter where
$M_2^*$= 1 $M_{\sun}$ and $r_p$ = 3.5 and c) an encounter where
$M_2^*$= 5 $M_{\sun}$ and $r_pd$ = 5.5. }
\label{fig:interact}
\end{figure}

However, it would be premature to declare encounters as not important
for the disc development. On the contrary, encounter-induced angular momentum 
transport may still play an important part in the disc evolution.
As one can see in Fig.~\ref{fig:interact}, an SAML
in the 3-5\% regime, which we showed to be present in the entire ONC, is 
sufficient to trigger pronounced spiral arm structures in the disc. As one can 
see similar structures appear independent of the mass of the perturber given 
the same angular momentum change.

As mentioned above such spiral structures have been detected in the few cases where discs 
have been observed with 
a sufficiently high spatial resolution \citep[for example][]{clampin:03,corder:05,lin:06}. 
A fly-by has often been rejected as possible cause of the spiral pattern because no stars have
been found in the immediate vicinity of the star-disc system.
However, as the results in Section 3 show, angular momentum change can be induced 
by relatively distant encounters given that the mass ratio between the star and the 
perturber is large. For example, for $M_2^*/M_1^*$=50 an encounter with a periastron
$\sim$14 times the disc radius could create a spiral arm pattern similar to those 
shown in  Fig.~\ref{fig:interact}. Considering the far range of interaction in many 
cases, a much larger area would have to be considered in a search for possible encounter 
partners.

What consequences can such spiral patterns in the disc have?
The spiral arm pattern above might be the starting point for gravitational 
instabilities (in above simulations self-gravity of the disc is not considered, so
clumping can not occur). So far gravitational instabilities have predominantly been 
investigated for isolated high-mass discs \citep[for a summary, see][]{durisen:06}. 
In this context gravitational instabilities can produce global spiral arms and disc 
fragmentation into dense clumps and substructures. Possibly these dense clumps may 
become precursors to gas giant planets \citep{kuiper:51,cameron:78,boss:98}. 

Modern numerical simulations, beginning with \cite{papaloizou:91}, show that 
non-axisymmetric disturbances
become unstable for $Q=c_s \kappa / \pi G \Sigma <$ 1.5, where $c_s$ is the
sound speed, $\kappa$ the epicyclic frequency, G the gravitational constant 
and $\Sigma$ the surface density. When the gravitational instabilities
emerge from the linear regime, they may either saturate at nonlinear 
amplitude or fragment the disc \citep{durisen:06}. What happens depends on 
whether a balance can be reached between heating and the loss of disk thermal 
energy by radiative or convective cooling. 
Although there is agreement on conditions for fragmentation, it is an open 
question whether real discs cool fast enough for fragmentation to occur and 
whether these fragments last long enough to contract into permanent 
protoplanets \citep{durisen:06}.

A prerequisite for the gravitational instabilities in isolated discs
is a high enough disc mass that self-gravity can carry small perturbations 
to fragmentation in the non-linear regime. However, observations show that
high-mass discs seem to be the exception rather than the rule.

\begin{figure}
\resizebox{\hsize}{!}{\includegraphics[angle=0]{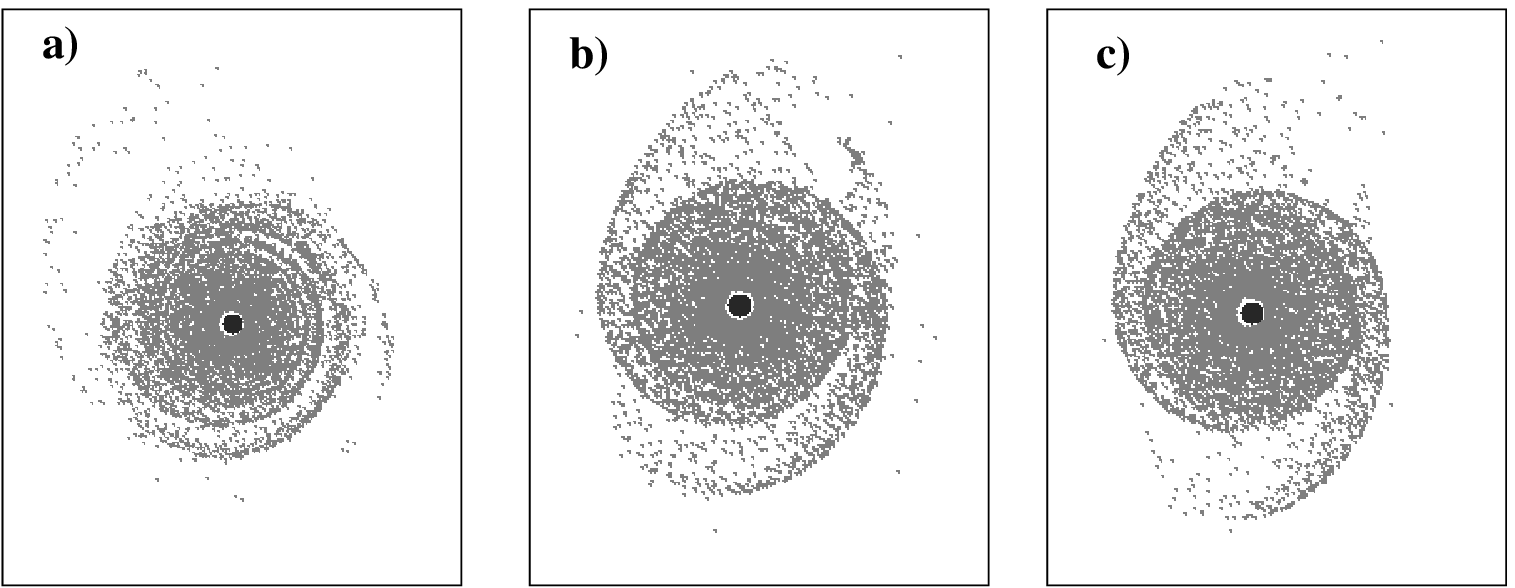}}
\caption{Examples of encounters with a SAML of $\sim$20\%. a) shows an encounter where
$M_2^*$= 0.1 $M_{\sun}$ and $r_p$ = 1.5, b)  an encounter where
$M_2^*$= 1 $M_{\sun}$ and $r_p$ = 2.8 and c) an encounter where
$M_2^*$= 5 $M_{\sun}$ and $r_p$ = 4.5. }
\label{fig:interact2}
\end{figure}

Various mechanisms have been suggested for triggering gravitational 
instabilities, among them formation of massive discs from the collapse of 
protostars \citep{laughlin:94,yorke:99,boss:98}, clumpy infall onto the disc
\citep{boss:98}, cooling from stable to unstable state, perturbation
by binary companions \citep{mayer:05} and close encounters with star/disc 
systems \citep{boffin:98,lin:98}. 

Investigating binary systems \cite{mayer:05} found that the degree of 
fragmentation depends on the mass of the disc and the cooling rate.
The simulations by Mayer et al.(2005) and Boss (2006) both indicate that
apparently, in contrast to isolated discs, in binary systems 
{\em lighter discs} are more likely to fragment. This counterintuitive result 
reflects the fact that intense heating from spiral shocks can suppress the 
formation of permanent clumps. A balance between compression/shock 
heating and cooling is required for fragmentation.

The situation in an encounter is similar to that in a binary system, because
the secondary star induces a non-axisymmetric perturbation onto the star-disc 
system, but with the difference that the perturbation is not recurring.
At the moment it is unclear whether encounters enhance disc fragmentation or not.
Investigations by \cite{lodato:06} have found that discs with long cooling timescales
are actually rendered more stable by encounters. On the other hand, \cite{shen:06} do 
get encounter induced fragmentation when they use an isothermal equation of state. 
In the future it needs to be determined by detailed simulations with
the right radiation hydrodynamics which amount of angular momentum change is 
required  to obtain stable clumps --- is it the stronger interactions 
(10-30\% angular momentum change) that we find in the cluster center or rather 
the weaker (3-5\% angular momentum change) ones further outside?

There is actually a second means by which spiral structures induced by encounters could support
planet formation. In the classical picture of planet \citep[for a summary see][]{weidenschilling:93} 
formation sticking of dust grains and
further agglomeration leads eventually to planet formation. Rice et al. 2006  showed that dust 
can accumulated in spiral shocks induced in self-gravitating disc. If this happens in the same 
fashion in encounter induced spirals, encounters could trigger planet formation via dust 
agglomeration as well. The higher dust density and possibly the higher velocity dispersion 
in the spiral arm could assist dust agglomeration. However, the time scale of the lifetime
of spiral structures is typically only 1000 to 2000 years which would be very stringent for 
this planet formation scenario. 
 
From the simulations in this paper we have shown that in systems like the ONC 
such an encounter situation is a common event throughout the cluster. 
Therefore, if planets can be formed through gravitational instabilities at all,
encounters in a cluster environment are a likely trigger for their occurance.




\section*{Acknowledgments}

We are grateful to R.Spurzem for providing the Nbody6++ code for the cluster 
simulations and want to thank the referee for the very helpful comments.
Simulations were partly performed at the John von 
Neumann Institute for Computing, Research Center J\"ulich, Project HKU14. 

\bibliographystyle{apj}

\clearpage
\begin{table}
\begin{scriptsize}
\begin{center}
\begin{tabular}{r||*{14}{r}}
 & 500.0 & 90.0 & 50.0 & 20.0 & 9.0 & 5.0 & 4.0 & 3.0 & 2.0 & 1.5 & 1.0 & 0.5 & 0.3 & 0.1\\[0.5ex]
\hline
\\[-2ex]
0.1 & 0.949 & 0.913 & 0.937 & 0.911 & 0.873& 0.797 & 0.713 & 0.747 & 0.736 & 0.708 & 0.688 & 0.633 & 0.488 & 0.239\\[0.5ex]
0.2 & 0.969 & 0.941 & 0.947 & 0.952 & 0.938& 0.900 & 0.863 & 0.866 & 0.860 & 0.826 & 0.719 & 0.592 & 0.391 & 0.212 \\[0.5ex]
0.3 & 0.990 & 0.991 & 0.986 & 0.969 & 0.959& 0.948 & 0.934 & 0.908 & 0.882 & 0.855 & 0.797 & 0.598 & 0.310 & 0.190 \\[0.5ex]
0.4 & 0.993 & 0.993 & 0.993 & 0.987 & 0.981& 0.963 & 0.953 & 0.935 & 0.891 & 0.897 & 0.784 & 0.577 & 0.238 & 0.170 \\[0.5ex]
0.5 & 0.996 & 0.994 & 0.993 & 0.992 & 0.970 & 0.949 & 0.941 & 0.919 & 0.894 & 0.856 & 0.783 & 0.554 & 0.201 & 0.164\\[0.05ex]
0.7 & 0.995 & 0.994 & 0.993 & 0.975 & 0.945 & 0.908 & 0.894 & 0.864 & 0.823 & 0.792 & 0.726 & 0.533 & 0.186& 0.143\\[0.5ex]
1.0 & 0.995 & 0.987 & 0.975 & 0.941 & 0.885 & 0.822 & 0.797 & 0.765 & 0.708 & 0.660 & 0.586 & 0.382 & 0.219 & 0.126\\[0.5ex]
1.5 & 0.992 & 0.955 & 0.926 & 0.860 & 0.762 & 0.681 & 0.642 & 0.587 & 0.501 & 0.447 & 0.340 & 0.200 & 0.152 & 0.048\\[0.5ex]
2.0 & 0.981 & 0.914 & 0.870 & 0.768 & 0.637 & 0.521 & 0.476 & 0.407 & 0.316 & 0.249 & 0.182 & 0.097 & 0.052 & 0.003\\[0.5ex]
2.5 & 0.963 & 0.865 & 0.805 & 0.662 & 0.508 & 0.375 & 0.321 & 0.263 & 0.187 & 0.136 & 0.079 & 0.021 & 0.006 & 0.000 \\[0.5ex]
3.0 & 0.943 & 0.810 & 0.728 & 0.560 & 0.380 & 0.239 & 0.198 & 0.143 & 0.089 & 0.060 & 0.020 & 0.003 & 0.001 & 0.000\\[0.5ex]
3.5 & 0.917 & 0.755 & 0.657 & 0.461 & 0.267 & 0.152 & 0.118 & 0.079 & 0.032 & 0.013 & 0.005 & 0.000 & 0.000 & 0.000 \\[0.5ex]
4.0 & 0.894 & 0.691 & 0.575 & 0.365 & 0.191 & 0.077 & 0.050 & 0.024 & 0.007 & 0.003 & 0.001 & 0.000 & 0.000 & 0.000 \\[0.5ex]
4.5 & 0.859 & 0.632 & 0.500 & 0.272 & 0.109 & 0.037 & 0.020 & 0.008 & 0.002 & 0.001 & 0.000 & 0.000 & 0.000 & 0.000 \\[0.5ex]
5.0 & 0.833 & 0.563 & 0.422 & 0.214 & 0.070 & 0.015 & 0.001 & 0.002 & 0.001 & 0.000 & 0.000 & 0.000 & 0.000 & 0.000 \\[0.5ex]
5.5 & 0.797 & 0.493 & 0.352 & 0.139 & 0.003 & 0.005 & 0.000 & 0.001 & 0.000 & 0.000 & 0.0& 0.0& 0.0& 0.0\\[0.5ex]
6.0 & 0.770 & 0.434 & 0.297 & 0.105 & 0.014 & 0.002 & 0.000 & 0.000 & 0.000 & 0.000 & 0.0& 0.0& 0.0& 0.0\\[0.5ex]
6.5 & 0.737 & 0.373 & 0.231 & 0.072 & 0.007 & 0.001 & 0.0& 0.000 & 0.0& 0.0& 0.0& 0.0& 0.0& 0.0\\[0.5ex]
7.0 & 0.703 & 0.313 & 0.198 & 0.036 & 0.003 & 0.001 & 0.0& 0.000 & 0.0& 0.0& 0.0& 0.0& 0.0& 0.0\\[0.5ex]
7.5 & 0.663 & 0.298 & 0.140 & 0.022 & 0.001 & 0.000 & 0.0& 0.0& 0.0& 0.0& 0.0& 0.0& 0.0& 0.0\\[0.5ex]
8.0 & 0.632 & 0.269 & 0.116 & 0.010 & 0.001 & 0.000 & 0.0& 0.0& 0.0& 0.0& 0.0& 0.0& 0.0& 0.0\\[0.5ex]
8.5 & 0.594 & 0.234 & 0.072 & 0.006 & 0.0 & 0.0& 0.0& 0.0& 0.0& 0.0& 0.0& 0.0& 0.0& 0.0\\[0.5ex]
9.0 & 0.557 & 0.209 & 0.056 & 0.003 & 0.0 & 0.0& 0.0& 0.0& 0.0& 0.0& 0.0& 0.0& 0.0& 0.0\\[0.5ex]
9.5 & 0.521 & 0.196 & 0.040 & 0.002 & 0.0& 0.0& 0.0& 0.0& 0.0& 0.0& 0.0& 0.0& 0.0& 0.0\\[0.5ex]
10.0& 0.479 & 0.162 & 0.022 & 0.001 & 0.0& 0.0& 0.0& 0.0& 0.0& 0.0& 0.0& 0.0& 0.0& 0.0\\[0.5ex]
10.5 & 0.437 & 0.117 & 0.015 & 0.001 & 0.0& 0.0& 0.0& 0.0& 0.0& 0.0& 0.0& 0.0& 0.0& 0.0\\[0.5ex]
11.0 & 0.406 & 0.107 & 0.008 & 0.001 & 0.0& 0.0& 0.0& 0.0& 0.0& 0.0& 0.0& 0.0& 0.0& 0.0\\[0.5ex]
11.5 & 0.375 & 0.127 & 0.006 & 0.000& 0.0& 0.0& 0.0& 0.0& 0.0& 0.0& 0.0& 0.0& 0.0& 0.0\\[0.5ex]
12.0 & 0.339 & 0.116 & 0.003 & 0.0& 0.0& 0.0& 0.0& 0.0& 0.0& 0.0& 0.0& 0.0& 0.0& 0.0\\[0.5ex]
12.5 & 0.316 & 0.097 & 0.002 & 0.0& 0.0& 0.0& 0.0& 0.0& 0.0& 0.0& 0.0& 0.0& 0.0& 0.0\\[0.5ex]
13.0 & 0.278 & 0.075 & 0.001 & 0.0& 0.0& 0.0& 0.0& 0.0& 0.0& 0.0& 0.0& 0.0& 0.0& 0.0\\[0.5ex]
13.5 & 0.266 & 0.053 & 0.001 & 0.0& 0.0& 0.0& 0.0& 0.0& 0.0& 0.0& 0.0& 0.0& 0.0& 0.0\\[0.5ex]
14.0 & 0.227 & 0.037 & 0.001 & 0.0& 0.0& 0.0& 0.0& 0.0& 0.0& 0.0& 0.0& 0.0& 0.0& 0.0\\[0.5ex]
14.5 & 0.219 & 0.026 & 0.001 & 0.0& 0.0& 0.0& 0.0& 0.0& 0.0& 0.0& 0.0& 0.0& 0.0& 0.0\\[0.5ex]
15.0 & 0.180 & 0.017 & 0.001 & 0.0& 0.0& 0.0& 0.0& 0.0& 0.0& 0.0& 0.0& 0.0& 0.0& 0.0\\[0.5ex]
15.5 & 0.176 & 0.013 & 0.0& 0.0& 0.0& 0.0& 0.0& 0.0& 0.0& 0.0& 0.0& 0.0& 0.0& 0.0\\[0.5ex]
16.0 & 0.170 & 0.011 & 0.0& 0.0& 0.0& 0.0& 0.0& 0.0& 0.0& 0.0& 0.0& 0.0& 0.0& 0.0\\[0.5ex]
16.5 & 0.137 & 0.011 & 0.0& 0.0& 0.0& 0.0& 0.0& 0.0& 0.0& 0.0& 0.0& 0.0& 0.0& 0.0\\[0.5ex]
17.0 & 0.129 & 0.009 & 0.0& 0.0& 0.0& 0.0& 0.0& 0.0& 0.0& 0.0& 0.0& 0.0& 0.0& 0.0\\[0.5ex]
17.5 & 0.099 & 0.0& 0.0& 0.0& 0.0& 0.0& 0.0& 0.0& 0.0& 0.0& 0.0& 0.0& 0.0& 0.0\\[0.5ex]
18.0 & 0.095 & 0.0& 0.0& 0.0& 0.0& 0.0& 0.0& 0.0& 0.0& 0.0& 0.0& 0.0& 0.0& 0.0\\[0.5ex]
18.5 & 0.070 & 0.0& 0.0& 0.0& 0.0& 0.0& 0.0& 0.0& 0.0& 0.0& 0.0& 0.0& 0.0& 0.0\\[0.5ex]
19.0 & 0.065 & 0.0& 0.0& 0.0& 0.0& 0.0& 0.0& 0.0& 0.0& 0.0& 0.0& 0.0& 0.0& 0.0\\[0.5ex]
19.5 & 0.046 & 0.0& 0.0& 0.0& 0.0& 0.0& 0.0& 0.0& 0.0& 0.0& 0.0& 0.0& 0.0& 0.0\\[0.5ex]
20.5 & 0.029 & 0.0& 0.0& 0.0& 0.0& 0.0& 0.0& 0.0& 0.0& 0.0& 0.0& 0.0& 0.0& 0.0\\[0.5ex]
21.  & 0.027 & 0.0& 0.0& 0.0& 0.0& 0.0& 0.0& 0.0& 0.0& 0.0& 0.0& 0.0& 0.0& 0.0\\[0.5ex]
21.5 & 0.025 & 0.0& 0.0& 0.0& 0.0& 0.0& 0.0& 0.0& 0.0& 0.0& 0.0& 0.0& 0.0& 0.0\\[0.5ex]
22.  & 0.017 & 0.0& 0.0& 0.0& 0.0& 0.0& 0.0& 0.0& 0.0& 0.0& 0.0& 0.0& 0.0& 0.0\\[0.5ex]
22.5 & 0.016 & 0.0& 0.0& 0.0& 0.0& 0.0& 0.0& 0.0& 0.0& 0.0& 0.0& 0.0& 0.0& 0.0\\[0.5ex]
23.  & 0.011 & 0.0& 0.0& 0.0& 0.0& 0.0& 0.0& 0.0& 0.0& 0.0& 0.0& 0.0& 0.0& 0.0\\[0.5ex]
23.5 & 0.010 & 0.0& 0.0& 0.0& 0.0& 0.0& 0.0& 0.0& 0.0& 0.0& 0.0& 0.0& 0.0& 0.0\\[0.5ex]
24.  & 0.006 & 0.0& 0.0& 0.0& 0.0& 0.0& 0.0& 0.0& 0.0& 0.0& 0.0& 0.0& 0.0& 0.0\\[0.5ex]
24.5 & 0.006 & 0.0& 0.0& 0.0& 0.0& 0.0& 0.0& 0.0& 0.0& 0.0& 0.0& 0.0& 0.0& 0.0\\[0.5ex]
25.  & 0.004 & 0.0& 0.0& 0.0& 0.0& 0.0& 0.0& 0.0& 0.0& 0.0& 0.0& 0.0& 0.0& 0.0\\[0.5ex]
26.  & 0.002 & 0.0& 0.0& 0.0& 0.0& 0.0& 0.0& 0.0& 0.0& 0.0& 0.0& 0.0& 0.0& 0.0\\[0.5ex]
27.  & 0.002 & 0.0& 0.0& 0.0& 0.0& 0.0& 0.0& 0.0& 0.0& 0.0& 0.0& 0.0& 0.0& 0.0\\[0.5ex]
28.  & 0.002 & 0.0& 0.0& 0.0& 0.0& 0.0& 0.0& 0.0& 0.0& 0.0& 0.0& 0.0& 0.0& 0.0\\[0.5ex]
29.  & 0.001 & 0.0& 0.0& 0.0& 0.0& 0.0& 0.0& 0.0& 0.0& 0.0& 0.0& 0.0& 0.0& 0.0\\[0.5ex]
30.0 & 0.001 & 0.0& 0.0& 0.0& 0.0& 0.0& 0.0& 0.0& 0.0& 0.0& 0.0& 0.0& 0.0& 0.0\\[0.5ex]
\end{tabular}
\caption{Table of AML $$(\Delta{J}_{\rm{d}}/J_{\rm{d}})_{bound}$$ for all simulated configurations of parabolic $(e=1)$ star-disc encounters. The first row contains the relative perturber masses $M_2^*/M_1^*$, the first column contains the relative periastra $r_{\rm{p}}/r_{\rm{d}}$. Results from simulations are denoted by four digits, the values ``0.0'' were edited manually.\label{table:jloss}}
\end{center}
\end{scriptsize}
\end{table}

\end{document}